\hfuzz 2pt
\vbadness 10000
\font\titlefont=cmbx10 scaled\magstep1

\magnification=\magstep1

\def\asymptotic#1\over#2{\mathrel{\mathop{\kern0pt #1}\limits_{#2}}}

\null
\vskip 1.5cm
\centerline{\titlefont DAMPED HARMONIC OSCILLATORS}
\medskip
\centerline{\titlefont IN THE HOLOMORPHIC REPRESENTATION}
\vskip 2.5cm
\centerline{\bf F. Benatti}
\smallskip
\centerline{Dipartimento di Fisica Teorica, Universit\`a di Trieste}
\centerline{Strada Costiera 11, 34014 Trieste, Italy}
\centerline{and}
\centerline{Istituto Nazionale di Fisica Nucleare, Sezione di
Trieste}
\vskip 1cm
\centerline{\bf R. Floreanini}
\smallskip
\centerline{Istituto Nazionale di Fisica Nucleare, Sezione di
Trieste}
\centerline{Dipartimento di Fisica Teorica, Universit\`a di Trieste}
\centerline{Strada Costiera 11, 34014 Trieste, Italy}
\vskip 2cm
\centerline{\bf Abstract}
\smallskip
\midinsert
\narrower\narrower\noindent
Quantum dynamical semigroups are applied to the study of the time evolution
of harmonic oscillators, both bosonic and fermionic. Explicit expressions
for the density matrices describing the states of these systems are
derived using the holomorphic representation. Bosonic and fermionic
degrees of freedom are then put together to form a supersymmetric
oscillator; the conditions that assure supersymmetry invariance of the
corresponding dynamical equations are explicitly derived.
\endinsert

\vfill\eject

{\bf 1. INTRODUCTION}
\bigskip

The dynamics of a small system ${\cal S}$ in interaction with a
large environment ${\cal E}$
is in general very complex and can not be described in terms of
evolution equations that are local in time. Possible initial correlations
and the continuous exchange of energy as well as entropy between the
${\cal S}$ and ${\cal E}$ produce phenomena of irreversibility and dissipation.

Nevertheless, there are instances for which simple and mathematically
precise description of the subdynamics can actually be given.
When the typical time scale in the evolution of the subsystem ${\cal S}$
is much larger than the characteristic time correlations in the
environment, one expects (and actually proves)
the disappearance of memory and non-linear
phenomena, although quantum coherence is usually lost.[1-5]

In such cases, the states of $\cal S$, conveniently described by
a density matrix $\rho$, are seen to evolve in time by means of a family
of linear maps that obey very basic physical requirements, like forward
in time composition (semigroup property) and complete positivity.
They form a so-called quantum dynamical semigroup.[1-3]

This description of the time evolution of open quantum systems is actually
very general; it is applicable to all physical situations for which
the interaction between ${\cal S}$ and ${\cal E}$ can be considered to be
weak and for times for which non-linear disturbances due to possible
initial correlations have disappeared. In particular, quantum dynamical
semigroups have been used to model laser dynamics in quantum optics,[6-8]
to study the evolution of various statistical systems,[1-3] to analyze
the interaction of a microsystem with a macroscopic apparatus.[9-11]

Recently, they have been used to describe effects leading to 
irreversibility and
dissipation in elementary particle physics phenomena.
Non-standard low energy effects accompanied by loss of quantum coherence
are in fact expected to appear as a consequence of gravitational quantum
fluctuations at Planck's scale.[12]
Detailed analysis of these effects have been performed in the system
of neutral mesons,[13-16] in neutron interferometry,[17]
neutrino oscillations,[18]
and in the propagation of polarized photons;[19, 20] the outcome of these
investigations is that present and future elementary particle experiments
will likely put stringent bounds on these non-standard dissipative phenomena.

These studies, in particular those
dealing with correlated neutral mesons,[14, 21]
have also further clarified the importance of the condition of complete
positivity in the description of open quantum systems. In many investigations
complete positivity is often replaced by the milder condition of simple
positivity; this guarantees the positivity of the eigenvalues of the
density matrix of the subsystem ${\cal S}$, but not that of a more
general system obtained by trivially coupling ${\cal S}$ with an arbitrary
$n$-level system. Lack of imposing this more stringent requirement
could lead to unacceptable physical consequences, like the appearance
of negative probabilities.[21]

To further analyze the properties of
the quantum dynamical semigroup description
of open systems, we shall apply this general framework
to the analysis of the evolution of one dimensional oscillators, both
bosonic and fermionic (for earlier investigations on the bosonic case,
see [22-25] and references therein).
We shall adopt the holomorphic representation [26, 27, 23]
since it allows an explicit description of the relevant density matrices
in terms of complex and Grassmannian (anticommuting) variables; in the most
simple situations, the general form of these density matrices
turns out to be Gaussian.
This allows the explicit evaluation of the corresponding (von Neumann) entropy
and analysis of its time evolution. Finally, we shall combine a bosonic
and a fermionic degrees of freedom to form a supersymmetric oscillator.
We shall then derive the conditions that guarantee the supersymmetric
invariance of the dynamical equations and discuss how these
affect the time evolution of the total density matrix.

\vskip 1cm

{\bf 2. THE BOSONIC OSCILLATOR}
\bigskip

As explained in the introductory remarks, we shall study the dynamics of a
single oscillator in interaction with a large environment. The states of the
system will be represented by a density matrix $\rho_B$, {\it i.e.} by
a positive hermitian operator, with constant trace, acting on the
bosonic Hilbert space ${\cal H}_B$. Our analysis is based on the
assumption that its time evolution is given by a quantum dynamical semigroup;
this is a completely positive, trace preserving, one parameter (=time)
family of linear maps, acting on the set $\{\rho\}$ of bosonic 
density matrices.
These maps are generated by equations of the following general form:[1-3]
$$
{\partial\rho(t)\over \partial t}={\cal L}[\rho(t)]\equiv
-i\big[H ,\rho(t)\big] + L[\rho(t)]\ .
\eqno(2.1)
$$
The first term in ${\cal L}$ is the standard quantum mechanical one,
that contains the system hamiltonian $H$, driving the time evolution
in absence of the environment. In the case of the bosonic oscillator,
it can be taken to have the most general quadratic form in the bosonic
creation $a^\dagger$ and annihilation $a$ operators:
$$
H_B={1\over2}\big[\omega_B\,( a^\dagger a +a a^\dagger)+\mu\, a^2
+\mu^*\, a^\dagger{}^2\big]\ ,
\eqno(2.2)
$$
where $\omega_B\geq0$ and $\mu$ is a complex parameter
(the star means complex conjugation).
The second piece $L[\rho]$ takes into account the interaction
with the environment; it is a linear map, whose form is fully determined
by the requirement of complete positivity and trace conservation:
$$
L[\rho]=-{1\over2}\sum_k\Big(L^\dagger_kL_k\,\rho +
\rho\, L^\dagger_kL_k\Big)\
+\sum_k L_k\,\rho\, L^\dagger_k\ .
\eqno(2.3)
$$
The operators $L_k$ should be chosen such that the expression in (2.3)
is well defined. In absence of the term $L[\rho]$, pure states would
be transformed into pure states. Instead, the additional piece (2.3) produce
in general dissipation and loss of quantum coherence.

The choice of the operators $L_k$ is largely arbitrary. However, since the
hamiltonian $H_B$ is quadratic in $a^\dagger$ and $a$, one is led to assume
the same property also for the additional term $L[\rho]$.
This implies a linear
expression for the operators $L_k$:
$$
L_k=r_k\, a + s_k\, a^\dagger\ ,
\eqno(2.4)
$$
with $r_k$ and $s_k$ complex parameters; this requirement further
guarantees the exact solvability of the equation in (2.1).
Note that the operators (2.4) are not bounded; nevertheless, by adapting the
arguments presented in Ref.[22] to the present case, one can show that
the exponential map generated by (2.1)
is well defined.

This description of the damped bosonic oscillator can be further simplified
by means of a suitable canonical transformation.
First, notice that not all values of the oscillator frequency $\omega_B$
and the complex parameter $\mu$ are physically allowed. Indeed, the spectrum
of the hamiltonian in (2.2) is bounded from below only for:
$$
\omega_B^2-|\mu|^2\geq0\ .
\eqno(2.5)
$$
This is a consequence of the fact that $H$ is an element of the Lie
algebra $su(1,1)$, whose generators in the so-called
metaplectic representation take the form:
$$
K_0={1\over4}(a^\dagger a+ a a^\dagger)\ ,\qquad
(K_+)^\dagger=K_-={a^2\over2}\ .
\eqno(2.6)
$$
The condition (2.5) guarantees that $H$ can be unitarily ``rotated''
to an element of the Cartan algebra with spectrum bounded from below.[23]

In other terms, by means of a unitary canonical transformation, one can now
pass to new operators:[26, 28]
$$
\eqalign{
&\tilde a=\Phi\, a +\Psi\, a^\dagger\ ,\cr
&\tilde a^\dagger=\Psi^*\,a+\Phi^*\, a^\dagger\ ,}
\eqno(2.7)
$$
with
$$
\Phi=\sqrt{\omega_B+\Omega_B\over 2\,\Omega_B }\ ,\qquad
\Psi={\mu^*\over\sqrt{2\,\Omega_B (\omega_B+\Omega_B)}}\ ,\qquad
\Omega_B=\sqrt{\omega_B^2-|\mu|^2}\ ,
\eqno(2.8)
$$
such that the hamiltonian take the simplified form:
$$
H_B={\Omega_B\over2}\,\{\tilde a^\dagger, \tilde a\}\ .
\eqno(2.9)
$$
The operators $L_k$ in (2.3) are still linear
in the new variables $\tilde a^\dagger$
and $\tilde a$, although with redefined coefficients.

This discussion explicitly shows that, without loss of generality, one can
set $\mu=\,0$ in (2.2); a non vanishing $\mu$ can always be reinstated at the
end by undoing the transformation (2.7).
With this choice, the evolution equation (2.1) for the bosonic oscillator
becomes:
$$
{\partial\rho_B(t)\over \partial t}={\cal L}_B
[\rho_B(t)]\equiv
-i\omega_B\big[a^\dagger a ,\rho_B(t)\big] + L_B[\rho_B(t)]\ ,
\eqno(2.10)
$$
where, by inserting (2.4) into (2.3), one has:
$$
\eqalign{
L_B[\rho]=\eta_B\Big(\big[a\rho, a^\dagger\big]+ \big[a,\rho 
a^\dagger\big]\Big)
&+\sigma_B\Big(\big[a^\dagger\rho, a\big]+ \big[a^\dagger,\rho a\big]\Big)\cr
&-\lambda^*_B \big[a,[a,\rho]\big]
-\lambda_B\big[a^\dagger, [a^\dagger,\rho]\big]\ ,}
\eqno(2.11)
$$
with
$$
\eta_B={1\over2}\sum_k|r_k|^2\ ,\qquad
\sigma_B={1\over2}\sum_k|s_k|^2\ ,\qquad
\lambda_B={1\over2}\sum_k r_k^* s_k\ .
\eqno(2.12)
$$
Note that from these expressions one deduces that:
$$
\eta_B\geq0\ ,\qquad \sigma_B\geq0\ ,\qquad
|\lambda_B|^2\leq \eta_B\sigma_B\ ,
\eqno(2.13)
$$
the last relation being a consequence of the Schwartz inequality;
let us remark that these are precisely the conditions that assure
complete positivity of the time evolution generated by the operator
$L_B$ in (2.11).

In order to study the solutions of the equation (2.10),
we shall work in the holomorphic representation;[26-28, 23]
it allows deriving explicit expressions for the density matrix $\rho_B(t)$
so that its behaviour in various regimes can be more easily discussed.
In this formulation, the elements $|\psi\rangle$ of the bosonic Hilbert
space ${\cal H}_B$ are represented by holomorphic functions $\psi(\bar z)$
of the complex variable $\bar z$, with inner product:%
\footnote{$^\dagger$}{Here and in the following we use the conventions
of Ref.[26]}
$$
\langle\phi|\psi\rangle=\int \psi^*(z)\, \phi(\bar z)\
e^{-\bar z z}\ d\bar z\, dz\ .
\eqno(2.14)
$$
To every operator $\cal O$ acting on ${\cal H}_B$ there correspond a kernel
${\cal O}(\bar z, z)$ of two independent complex variables $\bar z$ and $z$,
such that for the state $|\phi\rangle={\cal O}\,|\psi\rangle$ one finds
the representation:
$$
\phi(\bar z)=\int {\cal O}(\bar z, w)\, \psi(\bar w)\
e^{-\bar w w}\ d\bar w\, dw\ .
\eqno(2.15)
$$
In particular, the creation and annihilation operators, when acting on
a state $|\psi\rangle$, are realized by multiplication and
differentiation by the variable $\bar z$:
$$
a^\dagger |\psi\rangle \rightarrow \bar z\, \psi(\bar z)\ ,\qquad
a |\psi\rangle \rightarrow {\partial\over\partial\bar z} \psi(\bar z)\ ,
\eqno(2.16)
$$
while the identity operator is represented by $e^{\bar z  z}$.

Since the term ${\cal L}_B$ in (2.10) is at most quadratic
in $a^\dagger$ and $a$, the kernel $\rho_B(\bar z, z;t)$ representing the
solution of (2.10) can be taken to be of generic Gaussian form:
$$
\rho_B(\bar z, z;t)={1\over\sqrt{N(t)}}
e^{-{1\over2N(t)}\big[2\,y(t)\,\bar z z -\bar x(t)\, z^2 - x(t) {\bar z}^2\big]
+\bar z z}\ .
\eqno(2.17)
$$
Trace conservation for all times,
$$
{\rm Tr}[\rho_B(t)]=\int \rho_B(\bar z, z;t)\, e^{-\bar z z}\
d \bar z\, dz=1\ ,
\eqno(2.18)
$$
readily implies:
$$
N(t)=y^2(t)-|x(t)|^2\ ,
\eqno(2.19)
$$
while using (2.16) one finds that the unknown functions $x(t)$,
$\bar x(t)=[x(t)]^*$
and $y(t)$ have the following physical meaning:
$$
\eqalign{
&\langle a^2\rangle(t)\equiv {\rm Tr}\big[a^2\, \rho_B(t)\big]=
\int {\partial^2\over\partial{\bar z}^2}\big[\rho_B(\bar z, z;t)\big]\
e^{-\bar z z}\ d \bar z\, dz=x(t)\ ,\cr
&\langle a^\dagger{}^2\rangle(t)\equiv {\rm Tr}\big[a^\dagger{}^2\,
\rho_B(t)\big]=
\int {\bar z}^2\, \rho_B(\bar z, z;t)\
e^{-\bar z z}\ d \bar z\, dz=\bar x(t)\ ,\cr
&\langle a\, a^\dagger\rangle(t)
\equiv {\rm Tr}\big[a\, a^\dagger\,\rho_B(t)\big]=
\int {\partial\over\partial{\bar z}}\big[\bar z\, \rho_B(\bar z, z;t)\big]\
e^{-\bar z z}\ d \bar z\, dz=y(t)\ .}
\eqno(2.20)
$$
For simplicity, in writing (2.17) we have assumed
$\langle a^\dagger\rangle(t)=\langle a\rangle(t)=\,0$ for all times.
As shown in the Appendix, this
condition can be easily released starting with a more general Ansatz
for $\rho_B(\bar z, z;t)$; it will not be needed for the considerations
that follow.

Inserting (2.17) in the evolution equation (2.10), with the help of
the relations (2.16) one finds that the unknown functions
$x(t)$ and $y(t)$ satisfy the following linear equations:
$$
\eqalign{
&\dot x(t)=-2\,(\eta_B-\sigma_B+i\omega_B)\, x(t)-2\lambda_B\ ,\cr
&\dot y(t)=-2\,(\eta_B-\sigma_B)\, y(t)+2\eta_B\ .}
\eqno(2.21)
$$
General solutions can be easily obtained. For initial values
$x_0\equiv x(0)$, $y_0\equiv y(0)$ and $\eta_B\neq\sigma_B$, one finds:
$$
\eqalign{
&x(t)=E(t)\, e^{-2i\omega_B t}\, (x_0-x_\infty)+ x_\infty\ ,\cr
&y(t)=E(t)\, (y_0-y_\infty) + y_\infty\ ,}
\eqno(2.22)
$$
where
$$
E(t)=e^{-2(\eta_B-\sigma_B)t}\ ,\qquad
x_\infty={\lambda_B(\sigma_B-\eta_B+i\omega_B)\over
(\eta_B-\sigma_B)^2 + \omega_B^2}\ ,\qquad
y_\infty={\eta_B\over\eta_B-\sigma_B}\ ,
\eqno(2.23)
$$
while in the particular case $\eta_B=\sigma_B$:
$$
\eqalign{
&x(t)=e^{-2i\omega_B t}\, (x_0-x_c)+ x_c\ ,\qquad
x_c=i{\lambda_B\over\omega_B}\ ,\cr
&y(t)=2\eta_B t+y_0\ .}
\eqno(2.24)
$$

The large time behaviour of these solutions depends on the relative
magnitude of the two positive parameters $\eta_B$ and $\sigma_B$.
Only when $\eta_B>\sigma_B$, the functions $x(t)$ and $y(t)$ have a
well-defined limit. In this case, independently from the initial
conditions, the density matrix $\rho_B(t)$ approaches for large $t$
the equilibrium state $\rho_B^\infty$, obtained substituting in
(2.17) the asymptotic values $x_\infty$ and $y_\infty$ for $x(t)$ and $y(t)$.
Indeed, $\rho_B^\infty$ is clearly a fixed point of the evolution equation
(2.10), for any value of $\eta_B$ and $\sigma_B$.

Notice that $\rho_B^\infty$ does not correspond in general to a thermal
equilibrium state; to obtain an asymptotic Gibbs distribution, one has to
set $\lambda_B=\,0$ and introduce the inverse temperature $\beta$ via the
condition $y_\infty=[\coth(\beta\omega_B/2)+1]/2$
(compare with (2.27) and (2.28) below),
or equivalently $e^{\beta\omega_B}=\eta_B/\sigma_B$.

On the other hand, when $\eta_B<\sigma_B$, the exponential term $E(t)$
in (2.22) blows up for large times, while in the special case
$\eta_B=\sigma_B$, $y(t)$ grows linearly in time and $x(t)$ has an
oscillatory behaviour. In both cases, for generic initial conditions,
the normalization factor $N(t)$
in (2.19) grows unbounded, so that the functional
$\rho_B(\bar z, z;t)$ becomes vanishingly small, while retaining its
normalization, ${\rm Tr}[\rho_B(t)]=1$.%
\footnote{$^\dagger$}{For $\eta_B<\sigma_B$, the exception is given by
$\rho_B(t)\equiv\rho_B^\infty$, since, as noted before, $x_\infty$ 
and $y_\infty$
are fixed points of (2.21). Note that $y_\infty$ blows up for
$\eta_B=\sigma_B$, so that also $\rho_B^\infty$ becomes vanishingly
small in this limit. As a consequence, the infinite temperature limit
is singular.}

This peculiar behaviour can also be analyzed with the help of the Weyl
operators:
$$
W[\nu]=e^{\nu a+\bar\nu a^\dagger}\ .
\eqno(2.25)
$$
By studying the time evolution of $W$ induced by (2.10)
via the relation ${\rm Tr}[W(t)\,\rho_B]\equiv{\rm Tr}[W\,\rho_B(t)]$,
one finds that when $\eta_B>\sigma_B$ all Weyl operators remain well-defined
for all $t$, approaching the identity for large times; on the other hand,
for $\eta_B\leq\sigma_B$ one discovers that
all Weyl operators vanish in the large time limit,
except the identity $W[0]$, which is clearly
a fixed point of the time evolution.

As mentioned in the Introduction, the entropy of an open system
usually varies with time, due to the interaction with the
environment. In many physical instances, monotonic increase of the von Neumann
entropy,
$$
S[\rho]\equiv-{\rm Tr}[\rho\,\ln\rho]= -\langle \ln\rho\rangle\ ,
\eqno(2.26)
$$
is a desirable property.[1-3, 13-20]
For the damped oscillator described by the evolution
equation (2.10), this request can not be fulfilled in general, if one insists
on the existence of a well-behaved large-time equilibrium limit.

The explicit evaluation of $S_B\equiv S[\rho_B]$ is simplified by
noticing that to the representation kernel (2.17) there corresponds the
operator Ansatz:
$$
\rho_B=\bigg[{4\over\coth^2(\Omega/2)-1}\bigg]^{1/2}\
e^{-{1\over2}\big[A(a a^\dagger+a^\dagger a)+ B a^2 +\bar B a^\dagger{}^2\big]}
\ ,\eqno(2.27)
$$
where the parameters $A$ and $B$ are related to $x$ and $y$ of (2.17)
through the relations:
$$
x=-{B\over2\Omega}\coth{\Omega\over2}\ ,\qquad
y={1\over2}\bigg[{A\over\Omega}\coth{\Omega\over2}+1\bigg]\ ,\qquad
\Omega=\big(A^2-|B|^2\big)^{1/2}\ .
\eqno(2.28)
$$
Inserting these relations in the definition (2.26), one obtains:
$$
S_B=-{1\over2}\ln\bigg[{4\over\coth^2(\Omega/2)-1}\bigg]+
{1\over2}\Big[A\,\langle a a^\dagger+a^\dagger a\rangle+
B\, \langle a^2\rangle+ \bar B\, \langle a^\dagger{}^2\rangle\Big]\ .
\eqno(2.29)
$$
It is now convenient to introduce the following quantity:
$$
\eqalign{
\chi^2\equiv&{1\over4}\langle a a^\dagger+a^\dagger a\rangle^2
-\langle a^2\rangle\, \langle a^\dagger{}^2\rangle\cr
=&(y-1/2)^2-|x|^2={1\over4}\coth^2{\Omega\over2}\ ,}
\eqno(2.30)
$$
such that $\chi\geq 1/2$; in terms of this variable, one can easily rewrite
(2.29) as:
$$
S_B=\bigg(\chi+{1\over2}\bigg)\ln\bigg(\chi+{1\over2}\bigg)
-\bigg(\chi-{1\over2}\bigg)\ln\bigg(\chi-{1\over2}\bigg)\ .
\eqno(2.31)
$$

The entropy $S_B$ always grows with $\chi$, starting at the minimum
$S_B=\,0$ for $\chi=1/2$ and increasing as $\ln\chi$ for large $\chi$.
Recalling the explicit time-dependence of $x$ and $y$ in (2.22) and (2.24),
one realizes that in general for $\eta_B\neq\sigma_B$ the variable $\chi$
does not monotonically grows with $t$, so that the condition
$\dot S_B\geq0$ can not be satisfied for all times.
In particular, when $\eta_B>\sigma_B$ the equilibrium state $\rho_B^\infty$
is reached in general at the expense of some negative entropy-exchange
with the environment.

The case $\eta_B=\sigma_B$ is again special; in fact, the operators $L_k$
in (2.4) are now hermitian and therefore the condition $\dot S_B\geq0$ is
guaranteed. Indeed, in this case $\chi$ approaches infinity for large times,
and therefore so does $S_B$. Alternatively, using (2.3) in the definition
(2.26), one can directly show that:[29]
$$
\dot S_B\geq\Big\langle \sum_k \big[L_k,L_k^\dagger\big]\Big\rangle
=2(\sigma_B-\eta_B)\ .
\eqno(2.32)
$$

Note however that $L_k=L_k^\dagger$ is only a sufficient condition for
entropy increase. Indeed, for $\eta_B>\sigma_B$ take $x_0=x_\infty$
and $y_0\leq y_\infty$; in this case $\chi$ grows with time since
$\dot y(t)$ is always positive, and therefore also $S_B$ never decreases.

\vskip 1cm

{\bf 3. THE FERMIONIC OSCILLATOR}
\bigskip

We shall now extend the analysis of the previous section to the case
of a fermionic oscillator. The corresponding creation $\alpha^\dagger$
and annihilation $\alpha$ operators obey now the algebra:
$$
\{\alpha,\alpha^\dagger\}=1\ ,\qquad \alpha^2=\alpha^\dagger{}^2=\,0\ .
\eqno(3.1)
$$
As in the bosonic case, we shall assume the system in interaction with
a large environment, and describe its time evolution by means of a
quantum dynamical semigroup.

The states of the system will be described by an appropriate density
matrix $\rho_F$, acting on the elements of the fermionic Hilbert space
${\cal H}_F$. This operator obeys an evolution equation of the
form (2.1), where now the hamiltonian can be taken to be:
$$
H_F={\omega_F\over2}\,[\alpha^\dagger,\, \alpha]\ ,\qquad
\omega_F\geq0\ .
\eqno(3.2)
$$
Since $\alpha^\dagger$ and $\alpha$ are now nilpotent,
the additional piece $L[\rho]$ in (2.3) turns out to be
at most quadratic in these variables, and
the operators $L_k$ assume the generic form
$$
L_k=r_k^\prime\,\alpha + s_k^\prime\,\alpha^\dagger\ .
\eqno(3.3)
$$
Inserting this in (2.3), one explicitly finds:
$$
\eqalign{
L_F[\rho]=\eta_F\big(2\,\alpha\rho\alpha^\dagger-\alpha^\dagger\alpha\rho
-\rho\alpha^\dagger\alpha\big)
+\sigma_F\big(2\,&\alpha^\dagger\rho\alpha-\alpha\alpha^\dagger\rho
-\rho\alpha\alpha^\dagger\big)\cr
&+2\big(\lambda^*_F\,\alpha\rho\alpha
+\lambda_F\,\alpha^\dagger\rho\alpha^\dagger\big)\ ,}
\eqno(3.4)
$$
where the parameters $\eta_F$, $\sigma_F$ and $\lambda_F$
are as in (2.12) with the coefficients
$r_k$ and $s_k$ replaced by the primed ones.
Then, the complete evolution equation for the density matrix $\rho_F$
takes the form:
$$
{\partial\rho_F(t)\over \partial t}={\cal L}_F[\rho_F(t)]\equiv
-i\omega_F\big[\alpha^\dagger \alpha ,\rho_F(t)\big] + L_F[\rho_F(t)]\ .
\eqno(3.5)
$$

The study of the solutions of this equation in the holomorphic representation
requires the introduction of Grassmann variables
$\theta$, $\xi$, $\ldots$, that anticommute with the operators
$\alpha^\dagger$ and $\alpha$, and such that:
$$
\theta\xi=-\xi\theta\ ,\qquad \theta^2=\xi^2=\,0\ .
\eqno(3.6)
$$
The elements $|\psi\rangle$ of the Hilbert space ${\cal H}_F$
are now holomorphic functions $\psi(\bar\theta)$
of the variable $\bar\theta$. However, since ${\bar\theta}^2=\,0$,
their Taylor expansion contains only two terms:
$\psi(\bar\theta)=\psi_0 +\psi_1\, \bar\theta$, with $\psi_0$ and
$\psi_1$ complex parameters; they clearly represent the components
of $|\psi\rangle$ along the vacuum and one-fermion states.%
\footnote{$^\dagger$}{Since the fermionic oscillator is a two-level system,
a simple correspondence between the holomorphic and the
standard matrix representation can easily be established; however,
working with the holomorphic representation
is in general more convenient, since explicit, closed expressions
for $\rho_F$ can always be given, even in presence
of $n$ degrees of freedom. See also the discussion in Sect.5}

The inner product of two states $|\phi\rangle$ and $|\psi\rangle$
involves the integration over anticommuting variables (Berezin integral),
defined by the conditions $\int\theta\, d\theta=1$ and
$\int d\theta=\,0$:[26]
$$
\langle\phi|\psi\rangle=\int \psi^*(\theta)\,
\phi(\bar\theta)\ e^{-\bar\theta\theta}\ d\bar\theta\, d\theta\ ,
\eqno(3.7)
$$
where $\psi^*(\theta)=\psi_0^*+\theta\psi_1^*$ is by definition
the adjoint of $\psi(\bar\theta)$.

Similarly, to an operator $\cal O$ acting on ${\cal H}_F$ there corresponds
a kernel ${\cal O}(\bar\theta,\theta)$; the result of its action on the vector
$|\psi\rangle$ is given by:
$$
\phi(\bar\theta)=\int {\cal O}(\bar\theta,\xi)\,\psi(\bar\xi)\
e^{-\bar\xi\xi}\ d\bar\xi\, d\xi\ .
\eqno(3.8)
$$
Note that the identity operator is represented by the kernel
$e^{\bar\theta\theta}$. Furthermore, in this framework the fermionic
creation and annihilation operators are realized by left multiplication
and differentiation with respect to $\bar\theta$:
$$
\alpha^\dagger\rightarrow \bar\theta\ ,\qquad
\alpha\rightarrow{\partial\over\partial\bar\theta}\ ,
\eqno(3.9)
$$
so that $\alpha^\dagger$ is indeed the adjoint of $\alpha$ with respect to
the inner product in (3.7).

As in the bosonic case, since ${\cal L}_F$ in (3.5) is quadratic
in the operators (3.9), the kernel $\rho_F(\bar\theta,\theta;t)$
representing the state $\rho_F$ of the system can be taken to be
of Gaussian form:
$$
\rho_F(\bar\theta,\theta;t)=\gamma(t)\
e^{-\bar\theta\,{\mit\Gamma}(t)\,\theta}\ .
\eqno(3.10)
$$
For simplicity, also in this case we assume
$\langle\alpha^\dagger\rangle=\langle\alpha\rangle=\,0$ for all times,
so that terms linear in $\bar\theta$ and $\theta$ are absent in (3.10).
A more general Ansatz for $\rho_F(\bar\theta,\theta;t)$
is discussed in the Appendix.
The normalization condition:
$$
{\rm Tr}\big[\rho_F(t)\big]=
\int \rho_F(\bar\theta,\theta;t)\
e^{\bar\theta\theta}\ d\theta\, d\bar\theta=1\ ,
\eqno(3.11)
$$
readily implies: $\gamma(t)=\big[1-{\mit\Gamma}(t)\big]^{-1}$, so that
the kernel $\rho_F$ in (3.10) contains only one independent unknown function.
It can be conveniently recast in the following form:
$$
\rho_F(\bar\theta,\theta;t)=\gamma(t)+
\big[1-\gamma(t)\big]\,\bar\theta\theta\ ,
\eqno(3.12)
$$
explicitly showing that $\gamma$ and $1-\gamma$ represent the two eigenvalues
of $\rho_F$.%
\footnote{$^\dagger$}{Note that this simple rewriting
of the Gaussian Ansatz is possible only
in one dimension; in presence of $n$ degrees of freedom, the covariance
$\mit\Gamma$ would be an $n\times n$ hermitian
matrix and the Taylor expansion of (3.10)
would be much more involved.}
Finally, the physical meaning of $\gamma(t)$ can easily be derived:
$$
\langle\alpha\alpha^\dagger\rangle(t)\equiv
{\rm Tr}\big[\alpha\alpha^\dagger\,\rho_F(t)\big]
=\int {\partial\over\partial\bar\theta}
\Big[\bar\theta\, \rho_F(\bar\theta,\theta;t)\Big]\
e^{\bar\theta\theta}\ d\theta\, d\bar\theta=\gamma(t)\ .
\eqno(3.13)
$$

Insertion of (3.12) in the evolution equation (3.5) allows deriving
the equation satisfied by the unknown function $\gamma(t)$:
$$
\dot\gamma(t)=-2\,\big(\eta_F+\sigma_F\big)\,\gamma(t)+2\eta_F\ ,
\eqno(3.14)
$$
whose general solution is simply:
$$
\gamma(t)=e^{-2(\eta_F+\sigma_F)t} (\gamma_0-\gamma_\infty)+\gamma_\infty\ ,
\eqno(3.15)
$$
where $\gamma_0=\gamma(0)$ is the initial condition, while
$$
\gamma_\infty={\eta_F\over\eta_F+\sigma_F}\ .
\eqno(3.16)
$$
Since $\eta_F$ and $\sigma_F$ are positive constants, both
$\gamma(t)$ and $1-\gamma(t)$ are non negative, so that
$0\leq\gamma(t)\leq1$. Furthermore, independently from the initial condition,
the density matrix $\rho_F$ describing the
state of the fermionic oscillator always approaches for large times
the equilibrium configuration:
$\rho_F^\infty=\gamma_\infty+(1-\gamma_\infty)\bar\theta\theta$;
this is a thermal state, provided the inverse temperature $\beta$
is introduced via the relation: $e^{\beta\omega_F}=\eta_F/\sigma_F$,
with $\eta_F\geq\sigma_F$.

The evolution towards equilibrium is not in general associated with
a monotonic increase of the von Neumann entropy $S[\rho_F]\equiv S_F$.
Its explicit expression can be computed using the definition (2.26):
$$
S_F(t)=-\gamma(t)\ln\gamma(t)-\big[1-\gamma(t)\big]\ln\big[1-\gamma(t)\big]\ ,
\eqno(3.17)
$$
while its time derivative reads:
$\dot S_F=\dot\gamma[\ln(1-\gamma)-\ln\gamma]$; one can easily check
using (3.14) and (3.15) that $\dot S_F$ is always negative when $\gamma$
lays between $1/2$ and $\gamma_\infty$, while it is positive
outside this interval.

More precisely, as a function of $\gamma$, $S_F$ grows from its minimum
value $S_F=\,0$ at $\gamma=\,0$ up to its maximum $S_F=\ln2$ reached
for $\gamma=1/2$, and then decreases, becoming again zero at $\gamma=1$.
Therefore, $S_F$ monotonically grows only when $\gamma(t)$ increases in the
interval $[0,1/2]$, or decreases in the interval $[1/2,1]$.
For $\eta_F<\sigma_F$, this happens when $\gamma_0\leq\gamma_\infty$;
indeed, this implies $\gamma(t)\leq\gamma_\infty<1/2$ and $\dot\gamma(t)\geq0$
for all times. Similar conditions hold when $\eta_F>\sigma_F$;
in this case to obtain a monotonic increase of entropy, one has to choose
$\gamma_0\geq\gamma_\infty$, so that $\gamma(t)\geq\gamma_\infty>1/2$
and $\dot\gamma(t)\leq0$ for all $t$.

The case $\eta_F=\sigma_F$ is somehow special, since now $\dot S_F\geq0$
independently from the choice of the initial state;
the density matrix $\rho_F$ asymptotically
approaches the infinite-temperature, totally disordered state
$\rho_F^\infty=e^{\bar\theta\theta}/2$, for which the entropy is
maximal, \break $S_F=\ln2$.

As a final remark, note that in the case of the fermionic oscillator
the sufficient condition for entropy increase discussed at the end of the
previous section does not lead in general to useful constraints. Indeed,
the inequality in (2.32) gives now the condition:
$$
\dot S_F(t)\geq 2(\eta_F-\sigma_F)[1-2\,\gamma(t)]\ .
\eqno(3.18)
$$
Unless $\eta_F=\sigma_F$, the r.h.s. of this inequality becomes always
negative for large enough times, as it can be easily realized
by substituting for $\gamma(t)$ its asymptotic value $\gamma_\infty$.

\vskip 1cm

{\bf 4. THE SUPERSYMMETRIC OSCILLATOR}
\bigskip

We shall now discuss the behaviour of an oscillator composed by both bosonic
and fermionic degrees of freedom
in interaction with an environment, under
the hypothesis that its evolution is described by a quantum dynamical
semigroup. The density matrix $\rho$ representing the state of the system
is now an operator on the Hilbert space
${\cal H}={\cal H}_B\oplus{\cal H}_F$. Its time evolution is described
by an equation of the form (2.1), where both the total hamiltonian $H$
and the dissipative piece $L[\rho]$ are expressed in terms of bosonic,
$a^\dagger$, $a$, and fermionic, $\alpha^\dagger$, $\alpha$, creation and
annihilation operators, obeying
$$
[a^\dagger,\alpha^\dagger]=[a^\dagger,\alpha]
=[a,\alpha^\dagger]=[a,\alpha]=\,0\ ,
\eqno(4.1)
$$
together with the standard commutation, anticommutation relations.

The hamiltonian $H=H_B+H_F$, the sum of the bosonic and fermionic
terms of the form (2.9) and (3.2), possesses an additional property when the
two frequencies are equal: $\omega_B=\omega_F=\omega$.
Indeed, the following charges:
$$
Q_+=\omega^{1/2}\, a\,\alpha^\dagger\ ,\qquad
Q_-=\omega^{1/2}\, a^\dagger\alpha\ ,
\eqno(4.2)
$$
commute with the hamiltonian $H=\omega(a^\dagger a+\alpha^\dagger\alpha)$,
and further:
$$
\{Q_+,Q_-\}=H\ ,\qquad Q_+^2=Q_-^2=\,0\ .
\eqno(4.3)
$$
This is the simplest example of a supersymmetry algebra. The system
described by $H$ is therefore supersymmetric and
the conserved supercharges $Q_+$ and $Q_-$ exchange bosons and fermions;
further, from the algebra (4.3) one deduces that the ground state
of $H$ is a zero energy singlet and that
all excited states form degenerate doublets.

The additional piece $L[\rho]$ in the evolution equation (2.1)
will be taken to be the sum of the bosonic $L_B[\rho]$ and fermionic
$L_F[\rho]$ linear operators already introduced in the
previous sections. This is a natural choice since it assures integrability
of the time evolution ($L[\rho]$ is again at most quadratic in the creation
and annihilation operators), while avoiding mixings between
bosonic and fermionic degrees of freedom induced by the dissipative term;
in other terms, $L[\rho]$ is thus bosonic in character.

Nevertheless, this simple form of $L[\rho]$
does not in general assure supersymmetry
invariance. In ordinary Quantum Mechanics, to an invariance of the
hamiltonian there always correspond a conservation law and viceversa.
For time evolution generated by equations of the form (2.1) this is usually
not true: charge conservation and invariance (or symmetry) give rise to
two different and in general unrelated conditions.

To further elaborate on this point, notice that to the evolution equation
(2.1) for the density matrix $\rho$ there corresponds an analogous evolution
for any operator $X$ representing an observable of the system:
$$
{\partial\over\partial t}X={\cal L}^*[X]\equiv i[H,X]+L^*[X]\ ,
\eqno(4.4)
$$
where the linear operator $L^*$ is the ``dual'' of $L$ and
it is defined via the following identity
$$
{\rm Tr}\big(L^*[X]\ \rho\big)\equiv
{\rm Tr}\big(X\ L[\rho]\big)\ .
\eqno(4.5)
$$
Consider now a symmetry of the hamiltonian $H$ generated by the charge
$G$, inducing the following transformation on the observables:
$$
X\rightarrow X^\prime=U^{-1}\,X\,U\ ,\qquad U=e^{i\,G}\ .
\eqno(4.6)
$$
This transformation will be an invariance of the system only when it is
compatible with the evolution equation (4.4),
{\it i.e.} ${\cal L}^*[X^\prime]=U^{-1}\,{\cal L}^*[X]\,U$, for any $X$;
equivalently, in infinitesimal form:
$$
\big[G,L^*[X]\big]=L^*\big[ [G,X]\big]\ .
\eqno(4.7)
$$
This condition is clearly distinct from the relation that guarantees
the time conservation of the mean value
$\langle G\rangle\equiv {\rm Tr}[G\rho(t)]$ of the generator $G$;
recalling (2.1) and (4.5), from the condition $d/dt\langle G\rangle=\,0$
for any state, one readily derives:
$$
L^*[G]=\,0\ .
\eqno(4.8)
$$

In the case of the supersymmetric oscillator, the dual map
$L^*[X]=L^*_B[X]+L^*_F[X]$ can be easily deduced from (2.11) and (3.4).
Explicitly, one finds:
$$
\eqalign{
&L^*_B[X]=\eta_B\big(2\, a^\dagger X a- X a^\dagger a-a^\dagger a X\big)
+\sigma_B\big(2\, a X a^\dagger- X a a^\dagger-a a^\dagger X\big)\cr
&\hskip 2cm
+\lambda_B\big(2\, a^\dagger X a^\dagger- X a^\dagger{}^2-a^\dagger{}^2 X\big)
+\lambda^*_B\big(2\, a X a- X a^2-a^2 X\big)\ ,\cr
&L^*_F[X]=\eta_F\big(2\,\alpha^\dagger X\alpha-X\alpha^\dagger\alpha
-\alpha^\dagger\alpha X\big)
+\sigma_F\big(2\,\alpha X\alpha^\dagger-X\alpha\alpha^\dagger
-\alpha\alpha^\dagger X\big)\cr
&\hskip 2cm +2\,\lambda_F\,\alpha^\dagger X\alpha^\dagger
+2\,\lambda^*_F\,\alpha X\alpha
\ .}
\eqno(4.9)
$$
The parameter of a supersymmetry transformation is anticommuting,
so that the corresponding generator takes the form $G=\xi\, Q_+$, where
$\xi$ is a Grassmann variable, commuting with bosonic operators,
but anticommuting with the fermionic ones. Inserting it in (4.7)
and using (4.9), after some algebraic manipulations
one gets the following condition:
$$
\big(\eta_B-\sigma_B-\eta_F+\sigma_F\big)\,[X,G]+
\lambda^*_F\,\{X,G^\dagger\}=\,0\ .
\eqno(4.10)
$$
Since this relation must be true for any observable $X$,
supersymmetry invariance is compatible with the time evolution
only when:
$$
\eta_B-\sigma_B=\eta_F-\sigma_F\ ,\qquad \lambda_F=\,0\ .
\eqno(4.11)
$$

The holomorphic representation is again particularly
useful in order to discuss the behaviour of the state
$\rho(t)$ of the supersymmetric oscillator. The elements of the Hilbert space
$\cal H$ will be now represented by holomorphic functions
of the complex variable $\bar z$ and of the Grassmann symbol $\bar\theta$,
while creation and annihilation operators will act on them following
the rules in (2.16) and (3.9). The density matrix $\rho$ will be now
a kernel $\rho(\bar z, z;\bar\theta, \theta)$, whose explicit expression
can be taken to be of Gaussian form.%
\footnote{$^\dagger$}{Here again we assume vanishing initial averages
$\langle a^\dagger\rangle$, $\langle a\rangle$,
$\langle \alpha^\dagger\rangle$, $\langle \alpha\rangle$.}
It can be expanded as:
$$
\rho(\bar z, z;\bar\theta, \theta)=\rho_0(\bar z, z)+
\rho_1(\bar z, z)\ \bar\theta\theta\ .
\eqno(4.12)
$$
The normalization condition ${\rm Tr}[\rho]=1$ now involves both ordinary
and Grassmann integrals:
$$
\int \rho(\bar z, z;\bar\theta, \theta)\ e^{-\bar z z}\
e^{\theta\bar\theta}\ d\bar z\,dz\ d\theta\,d\bar\theta=
\int [\rho_0(\bar z, z)+\rho_1(\bar z, z)]\, e^{-\bar z z}\
d\bar z\,dz=1\ .
\eqno(4.13)
$$

Inserting the Ansatz (4.12) into the evolution equation for $\rho$ allows
deriving the following conditions on the bosonic kernels
$\rho_0$ and $\rho_1$:
$$
\eqalignno{
&\dot\rho_0(t)={\cal L}_B[\rho_0(t)]
+2\,\big[\eta_F\,\rho_1(t)-\sigma_F\,\rho_0(t)\big]\ , &(4.14a)\cr
&\dot\rho_1(t)={\cal L}_B[\rho_1(t)]
-2\,\big[\eta_F\,\rho_1(t)-\sigma_F\,\rho_0(t)\big]\ , &(4.14b)\cr}
$$
where the linear operator ${\cal L}_B[\rho]$ is as in (2.10).
It follows that the combination $\rho_0+\rho_1$ satisfies the same
evolution equation discussed in Sect.2 for the case of a single
bosonic oscillator. The Gaussian Ansatz $\rho_B(\bar z, z)$ in (2.17)
can equally well be adopted here for $\rho_0+\rho_1$, since performing
the Grassmann integrations, one consistently finds (compare with (2.20)):
$$
\eqalign{
&\langle a^2\rangle(t)\equiv {\rm Tr}\big[a^2\, \rho(t)\big]=
\int {\partial^2\over\partial{\bar z}^2}
\big[\rho_0(\bar z, z;t)+\rho_1(\bar z, z;t)\big]\
e^{-\bar z z}\ d \bar z\, dz=x(t)\ ,\cr
&\langle a^\dagger{}^2\rangle(t)\equiv {\rm Tr}\big[a^\dagger{}^2\,
\rho(t)\big]=
\int {\bar z}^2\, \big[\rho_0(\bar z, z;t)+\rho_1(\bar z, z;t)\big]\
e^{-\bar z z}\ d \bar z\, dz=\bar x(t)\ ,\cr
&\langle a\, a^\dagger\rangle(t)
\equiv {\rm Tr}\big[a\, a^\dagger\,\rho(t)\big]=
\int {\partial\over\partial{\bar z}}\bar z
\big[\rho_0(\bar z, z;t)+\rho_1(\bar z, z;t)\big]\
e^{-\bar z z}\ d \bar z\, dz=y(t)\ .}
\eqno(4.15)
$$
As a consequence, the time evolution of these quantities is that
given in (2.22) and (2.24).

Inserting back this result into $(4.14a)$, one obtains:
$$
\dot\rho_0(t)={\cal L}_B[\rho_0(t)]
-2\,\big(\eta_F+\sigma_F\big)\rho_0(t)+2\eta_F\,\rho_B(t)\ .
\eqno(4.16)
$$
The form of this equation suggests to look for a solution
in which $\rho_0(t)$ differs from $\rho_B(t)$ by an unknown multiplicative
function $\gamma_F(t)$. It can be identified with the function
$\gamma(t)$ studied in the previous section,
since it satisfies the same equation (3.14) and
has the same physical meaning:
$$
\langle\alpha\alpha^\dagger\rangle(t)\equiv
{\rm Tr}\big[\alpha\alpha^\dagger\,\rho(t)\big]
=\int \rho_0(\bar z,z;t)\ e^{-z\bar z}\ d z\, d\bar z=\gamma_F(t)\ .
\eqno(4.17)
$$
As a consequence, $\rho_1=(1-\gamma_F)\rho_B$, and therefore one
finally finds:
$$
\rho(\bar z, z;\bar\theta, \theta)=
\Big[\gamma_F+(1-\gamma_F)\,\theta\bar\theta\Big]\, \rho_B(\bar z, z)
\equiv\rho_F(\bar\theta, \theta)\ \rho_B(\bar z, z)\ .
\eqno(4.18)
$$

Not surprisingly, the density matrix that solves the evolution
equation (2.1) in the case of the supersymmetric oscillator
is in factorized form; its behaviour can be deduced
from the analysis of the previous sections,
provided the conditions (4.11) for supersymmetry invariance are taken
into account.

In particular, $\rho$ approaches an equilibrium state for large times
only when $\eta_B>\sigma_B$, which also implies: $\eta_F>\sigma_F$.
This limiting state is thermal, with inverse temperature $\beta$,
only for $\lambda_B=\,0$ and
$\eta_B/\sigma_B=\eta_F/\sigma_F=e^{\beta\omega}$,
which implies, recalling the condition (4.11):
$\eta_B=\eta_F$ and $\sigma_B=\sigma_F$.[30]

Also in the case of the supersymmetric oscillator, the total entropy
does not have in general a monotonic behaviour during the approach to
equilibrium. Since the density matrix $\rho$ is in factorized form,
the total entropy $S$ will be the sum of the bosonic and fermionic
contributions.
Using the variable $\gamma_B=\chi-1/2\geq0$, where $\chi$
is defined as in (2.30), and
recalling the results of the previous sections,
one has:
$$
S=\big(\gamma_B+1\big)\ln\big(\gamma_B+1\big)-\gamma_B\ln\gamma_B
-\big(\gamma_F-1\big)\ln\big(\gamma_F-1\big)-\gamma_F\ln\gamma_F\ .
\eqno(4.19)
$$
Its time derivative, that can be expressed as:
$$
\dot S=\dot\gamma_B\ln\bigg(1+{1\over\gamma_B}\bigg)
+\dot\gamma_F\ln\bigg(1-{1\over\gamma_F}\bigg)\ ,
\eqno(4.20)
$$
does not have in general a definite sign,
although possible compensations between the bosonic and fermionic
contributions can concur to a positive r.h.s. for certain time intervals.

As discussed at the end of Sect.2, a bound on $\dot S$ can be
obtained by working directly with the definition (2.26) and
the equation (2.1). In the present case, this procedure gives:
$$
\dot S\geq 2\big(\sigma_B-\eta_B\big)
+2\big(\sigma_F-\eta_F\big)\big[2\,\gamma_F-1\big]\equiv
4\big(\sigma_F-\eta_F\big)\, \gamma_F\ ,
\eqno(4.21)
$$
where the identity is a consequence of the condition (4.11).
Since $0\leq\gamma_F(t)\leq1$, the inequality (4.21) assures
$\dot S\geq0$ for $\sigma_F\geq\eta_F$. However, this condition
would lead to a rather singular behaviour for the bosonic part
of the density matrix in (4.18), and thus for the whole $\rho$.
In fact, also $\sigma_B$ would be greater than $\eta_B$ and,
as discussed in Sect.2,
this implies an infinitely growing average occupation number.
In conclusion, although inducing a
partial compensation between the bosonic and fermionic contributions
to $S$, the supersymmetry condition (4.11)
is in general not enough to guarantee monotonic entropy
increase for all times during the evolution of the system.

\vskip 1cm

{\bf 5. DISCUSSION}
\bigskip

All the considerations developed in the
previous sections for single oscillators can be
generalized to the case of $n$ independent oscillators,
both bosonic and fermionic.
Their interaction with an external environment can still be
consistently described in terms of quantum dynamical semigroups,
so that their time evolution can be modelled by means of equations of
the form (2.1), (2.3), with operators $L_k$ linear in the relevant
fundamental variables. However, the coefficients $r$ and $s$
of (2.4) become now matrices, and the number of independent constants
characterizing the dissipative part $L[\rho]$ rapidly increase
with $n$, making the evolution equation (2.1) rather involved.

Nevertheless, various simplifying conditions
can be imposed to reduce, at least in part this arbitrariness.
Those involving symmetry properties are the most physically interesting.
As discussed in the Introduction,
the interaction between system
and environment can be considered in general to be weak; therefore,
in many instances, the presence of the environment should
not to be able to alter the symmetry properties of the system.
In the case of $n$ isotropic oscillators,
the hamiltonian is invariant under the action of the group $SU(n)$;
it is then quite natural to assume the same invariance property
to be valid for the full evolution equation. As discussed in Sect.4,
this can be achieved by imposing the condition (4.7) for any
element $G$ of the $SU(n)$ algebra.

Also in this more general setting,
the holomorphic representation appears to be
a particularly convenient framework to analyze the behaviour
of the solutions of (2.1).
It requires the introduction of
$n$ commuting or anticommuting complex symbols, that allow realizing
the corresponding creation and annihilation operators
as multiplication and differentiation by these variables.
The kernel representing the system density matrix
can still be taken to have a generic Gaussian functional expression.
However, the various ``coefficients'' in the exponent, suitable generalizations
of the functions $x$, $\bar x$ and $y$ in the bosonic case and of
$\mit\Gamma$ in the fermionic one, become now $n\times n$ matrices.
They obey quadratic (Riccati-like) time evolution equations,
whose solutions can always be obtained,
albeit in general in terms of implicitly defined functions.[31, 32]

Although developed in the analysis of simple open systems,
the techniques described in the previous sections are actually
very general; they can be used
to study the dynamics of more complicated models,
for which the operator $\cal L$
in (2.1) is not quadratic in the relevant variables.
In these cases, complete explicit expressions for the density
matrix $\rho$ as solution of (2.1) can not in general be given.
Nevertheless, approximate expressions for $\rho$, typically in Gaussian
form, can be obtained via the application of suitable variational procedures.

Indeed, equations of the form (2.1) can be derived by mean of
a suitable variational principle,[33] obtained by generalizing the one yielding
the Liouville - von Neumann equation in ordinary Quantum Mechanics.[28, 34]
In the case of isoentropic time evolutions, these variational techniques
have allowed detailed
discussions of a wide range of physical phenomena, from
statistical physics to inflationary cosmology.[34, 35]
Their application to the study of
quantum dynamical semigroups within the framework presented
in the previous sections
will surely provide new insights on the behaviour
of open quantum systems.

\vskip 1cm

{\bf APPENDIX}
\bigskip

The Gaussian kernels $\rho_B(\bar z, z;t)$ and
$\rho_F(\bar\theta, \theta;t)$ representing the density matrices
for the bosonic and fermionic oscillators discussed in
Sect.2 and 3 lead to vanishing averages for the corresponding
creation and annihilation operators. This condition can easily
be released by introducing a more general Ansatz.

In the bosonic case, take:
$$
\rho_B(\bar z, z;t)={1\over\sqrt{N(t)}}
e^{-{1\over2N(t)}\big[2\,y(t)\,[\bar z-\bar v(t)][z-v(t)]
-\bar x(t)\, [z-v(t)]^2 - x(t)[\bar z-\bar v(t)]^2\big]
+\bar z z}\ ,
\eqno(A.1)
$$
that differ from the expression (2.17) because of the presence
of the two additional functions $v(t)$ and $\bar v(t)$.
Trace conservation for all times,
${\rm Tr}[\rho_B(t)]=1$, still implies:
$$
N(t)=y^2(t)-|x(t)|^2\ ,
\eqno(A.2)
$$
while hermiticity requires: $\bar v(t)\equiv [v(t)]^*$.
With this choice for $\rho_B(\bar z, z;t)$, the averages of
$a^\dagger$ and $a$ are in general non vanishing:
$$
\eqalign{
&\langle a^\dagger\rangle(t)\equiv {\rm Tr}\big[a^\dagger\, \rho_B(t)\big]=
\int \bar z\,\rho_B(\bar z, z;t)\
e^{-\bar z z}\ d \bar z\, dz=\bar v(t)\ ,\cr
&\langle a\rangle(t)\equiv {\rm Tr}\big[a\, \rho_B(t)\big]=
\int {\partial\over\partial{\bar z}}\big[\rho_B(\bar z, z;t)\big]\
e^{-\bar z z}\ d \bar z\, dz=v(t)\ .}
\eqno(A.3)
$$
The time evolution equation (2.10) implies the following homogeneous
equation for $v(t)$
$$
\dot v(t)=2\,(\eta_B-\sigma_B+i\omega_B)\, v(t)\ ,
\eqno(A.4)
$$
so that $v(t)$ is non vanishing only if its initial value
$v(0)$ is different from zero:
$$
v(t)=e^{-(\eta_B-\sigma_B+i\omega_B)t}\ v(0)\ .
\eqno(A.5)
$$

The physical meaning of the remaining functions $x$, $\bar x$ and $y$
appearing in $(A.1)$ is slightly changed with respect to those
studied in Sect.2; now one finds:
$$
\eqalign{
&x=\langle a^2\rangle- \langle a\rangle^2\ ,\cr
&\bar x=\langle a^\dagger{}^2\rangle- \langle a^\dagger\rangle^2\ ,\cr
&y=\langle a\,a^\dagger\rangle- \langle a\rangle\,\langle a^\dagger\rangle\ .}
\eqno(A.6)
$$
Nevertheless, one can check that these functions
still obey the evolution equations (2.21),
so that the considerations and the discussions of Sect.2 apply
to this more general situation as well.

In the case of the fermionic oscillator, the most general Gaussian Ansatz
for the kernel $\rho_F(\bar\theta, \theta;t)$ can be written as:
$$
\rho_F(\bar\theta,\theta;t)=\gamma(t)\
e^{-{1\over\gamma(t)}\big[\bar\theta\,{\mit\Delta}(t)\,\theta
-\bar\varphi(t)\,\theta-\varphi(t)\,\bar\theta\big]}\ .
\eqno(A.7)
$$
The normalization condition ${\rm Tr}\big[\rho_F(t)\big]=1$
gives ${\mit\Delta(t)}=\gamma(t)-1$, while hermiticity implies
$\bar\varphi(t)=[\varphi(t)]^*$.

By performing the integration over the anticommuting variables,
one finds that the function $\gamma(t)$ retains its meaning
as $\langle\alpha\, \alpha^\dagger\rangle$ also in this more general
setting, and therefore still obeys the evolution equation (3.14).

On the other hand, the two additional functions $\bar\varphi(t)$ and
$\varphi(t)$ in $(A.7)$ represent the averages of
$\alpha^\dagger$ and $\alpha$,
$$
\eqalign{
&\langle\alpha^\dagger\rangle(t)\equiv{\rm Tr}[\alpha^\dagger\, \rho_F(t)]=
\int\bar\theta\,\rho_F(\bar\theta,\theta;t)\
e^{\bar\theta\theta}\ d\theta\, d\bar\theta=\bar\varphi(t)\ ,\cr
&\langle\alpha\rangle(t)\equiv{\rm Tr}[\alpha\, \rho_F(t)]=
\int {\partial\over\partial\bar\theta}
\big[\rho_F(\bar\theta,\theta;t)\big]\
e^{\bar\theta\theta}\ d\theta\, d\bar\theta=\varphi(t)\ ,}
\eqno(A.8)
$$
and, as a consequence of (3.5), obey the following evolution equations:
$$
\eqalign{
&\dot\varphi(t)=-2\,(\eta_F+\sigma_F+i\omega_F)\,\varphi(t)
+2\lambda_F\,\bar\varphi(t)\ ,\cr
&\dot{\bar\varphi}(t)=-2\,(\eta_F+\sigma_F-i\omega_F)\,\bar\varphi(t)
+2\lambda^*_F\,\varphi(t)\ .}
\eqno(A.9)
$$
The general solution is given by:
$$
\varphi(t)=e^{-(\eta_F+\sigma_F)t}\Bigg\{
\bigg[\cos(\Omega_F t)-{i\omega_F\over\Omega_F}\,
\sin(\Omega_F t)\bigg]\,\varphi(0)
+{2\,\lambda_F\over\Omega_F}\,\sin(\Omega_F t)\, \bar\varphi(0)\Bigg\}\ ,
\eqno(A.10)
$$
where $\Omega_F=\big[\omega_F^2-4|\lambda_F|^2\big]^{1/2}$
for $\omega_F\geq|\lambda_F|$.
Hyperbolic functions appear in the expression $(A.10)$ when
$\omega_F<|\lambda_F|$; however, thanks to the inequality
$|\lambda_F|^2\leq\eta_F\sigma_F$ (compare with (2.13)), $\varphi(t)$ always
vanishes for large times.

\vfill\eject

{\bf REFERENCES}
\bigskip

\item{1.} R. Alicki and K. Lendi, {\it Quantum Dynamical Semigroups and
Applications}, Lect. Notes Phys. {\bf 286}, (Springer-Verlag, Berlin, 1987)
\smallskip
\item{2.} V. Gorini, A. Frigerio, M. Verri, A. Kossakowski and
E.C.G. Surdarshan, Rep. Math. Phys. {\bf 13} (1978) 149
\smallskip
\item{3.} H. Spohn, Rev. Mod. Phys. {\bf 53} (1980) 569
\smallskip
\item{4.} A. Royer, Phys. Rev. Lett. {\bf 77} (1996) 3272
\smallskip
\item{5.} F. Benatti and R. Floreanini, Ann. of Phys. {\bf 273} (1999) 58
\smallskip
\item{6.} W.H. Louisell, {\it Quantum Statistical Properties of Radiation},
(Wiley, New York, 1973)
\smallskip
\item{7.} M.O. Scully and M.S. Zubairy,
{\it Quantum Optics} (Cambridge University Press, Cambridge, 1997)
\smallskip
\item{8.} C.W. Gardiner and P. Zoller,
{\it Quantum Noise}, 2nd. ed. (Springer, Berlin, 2000)
\smallskip
\item{9.} L. Fonda, G.C. Ghirardi and A. Rimini, Rep. Prog. Phys.
{\bf 41} (1978) 587
\smallskip
\item{10.} H. Nakazato, M. Namiki and S. Pascazio,
Int. J. Mod. Phys. {\bf B10} (1996) 247
\smallskip
\item{11.} F. Benatti and R. Floreanini, Phys. Lett. {\bf B428} (1998) 149
\smallskip
\item{12.} S. Hawking, Comm. Math. Phys. {\bf 87} (1983) 395; Phys. Rev. D
{\bf 37} (1988) 904; Phys. Rev. D {\bf 53} (1996) 3099;
J. Ellis, J.S. Hagelin, D.V. Nanopoulos and M. Srednicki,
Nucl. Phys. {\bf B241} (1984) 381
\smallskip
\item{13.} F. Benatti and R. Floreanini, Nucl. Phys. {\bf B488} (1997) 335
\smallskip
\item{14.} F. Benatti and R. Floreanini, Nucl. Phys. {\bf B511} (1998) 550
\smallskip
\item{15.} F. Benatti and R. Floreanini, Phys. Lett. {\bf B465}
(1999) 260
\smallskip
\item{16.} F. Benatti and R. Floreanini, $CPT$, dissipation, and all that,
in the {\it Proceedings of Daphne99}, Frascati, November 1999,
{\tt hep-ph/9912426}
\smallskip
\item{17.} F. Benatti and R. Floreanini, Phys. Lett. {\bf B451} (1999) 422
\smallskip
\item{18.} F. Benatti and R. Floreanini, JHEP {\bf 02} (2000) 032
\smallskip
\item{19.} F. Benatti and R. Floreanini, Effective dissipative dynamics
for polarized photons, Trieste-preprint, 2000
\smallskip
\item{20.} F. Benatti, R. Floreanini and A. Lapel, Open quantum systems
and complete positivity, Trieste-preprint, 2000
\smallskip
\item{21.} F. Benatti and R. Floreanini,
Mod. Phys. Lett. {\bf A12} (1997) 1465;
Banach Center Publications, {\bf 43} (1998) 71;
Phys. Lett. {\bf B468} (1999) 287; On the weak-coupling limit and complete
positivity, Chaos Sol. Frac., to appear
\smallskip
\item{22.} G. Lindblad, Rep. Math. Phys. {\bf 10} (1976) 393
\smallskip
\item{23.} A. Perelomov, {\it Generalized Coherent States
and Their Applications}, (Springer-Verlag, Berlin, 1986)
\smallskip
\item{24.} A. Sandulescu and H. Scutaru, Ann. of Phys. {\bf 173} (1987) 277
\smallskip
\item{25.} A. Isar, Fortschr. Phys. {\bf 47} (1999) 855
\smallskip
\item{26.} F.A. Berezin, {\it The Method of Second Quantization},
(Academic Press, Orlando, 1966)
\smallskip
\item{27.} L.D. Faddeev, Introduction to functional methods, in {\it Methods
in Field Theory}, R. Balian, J. Zinn-Justin, eds., (North-Holland, Amsterdam,
1976)
\smallskip
\item{28.} J.-P. Blaizot and G. Ripka, {\it Quantum Theory of Finite Systems},
(The MIT Press, Cambridge, 1986)
\smallskip
\item{29.} F. Benatti and H. Narnhofer, Lett. Math. Phys. {\bf 15} (1988) 325
\smallskip
\item{30.} L. Van Hove, Nucl. Phys. {\bf B207} (1982) 15
\smallskip
\item{31.} R. Floreanini, Ann. of Phys. {\bf 178} (1987) 227
\smallskip
\item{32.} R. Floreanini and R. Jackiw, Phys. Rev. D {\bf 37} (1988) 2206
\smallskip
\item{33.} A.K. Rajagopal, Phys. Lett. {\bf A228} (1997) 66
\smallskip
\item{34.} O. Eboli, R. Jackiw and S.-Y. Pi, Phys. Rev. D {\bf 37} (1988) 3557
\smallskip
\item{35.} O. Eboli, S.-Y. Pi and M. Samiullah, Ann. of Phys.
{\bf 193} (1989) 102

\bye